\documentclass[twocolumn,showpacs,preprintnumbers,amsmath,amssymb]{revtex4}%
\usepackage{graphicx}
\usepackage{dcolumn}
\usepackage{bm}
\usepackage{color}
\usepackage{amsmath}
\usepackage{amsfonts}
\usepackage{amssymb}
\usepackage{ulem}
\begin{document}
\preprint{APS/123-QED}
\title{Recollision of excited electron in below-threshold nonsequential double ionization}
\author{Xiaolei Hao,$^{1}$}
\author{Yuxing Bai$^{1}$}
\author{Chan Li$^{1}$}
\author{Jingyu Zhang$^{1}$}
\author{Weidong Li$^{1}$}
\author{Weifeng Yang$^{2}$}
\email{wfyang@stu.edu.cn}
\author{MingQing Liu$^{3}$}
\author{Jing Chen$^{3,4}$}
\email{chen_jing@iapcm.ac.cn} \affiliation{$^{1}$Institute of
Theoretical Physics and Department of Physics, State Key
Laboratory of Quantum Optics and Quantum Optics Devices,
Collaborative Innovation Center of Extreme Optics, Shanxi
University, Taiyuan 030006, China} \affiliation{$^{2}$Department
of Physics, College of Science, Shantou University, Shantou,
Guangdong 515063, China} \affiliation{$^{3}$Institute of Applied
Physics and Computational Mathematics, P. O. Box 8009, Beijing
100088, China} \affiliation{$^{4}$Shenzhen Key Laboratory of
Ultraintense Laser and Advanced Material Technology, Center for
Advanced Material Diagnostic Technology, and College of
Engineering Physics, Shenzhen Technology University, Shenzhen
518118, China}
\date{\today}

\begin{abstract}
Consensus has been reached that recollision, as the most important
post-tunneling process, is responsible for nonsequential double
ionization process in intense infrared laser field, however, its
effect has been restricted to interaction between the first
ionized electron and the residual univalent ion so far. Here we
identify the key role of recollision between the second ionized
electron and the divalent ion in the below-threshold nonsequential
double ionization process by introducing a Coulomb-corrected
quantum-trajectories method, which enables us to well reproduce
the experimentally observed cross-shaped and anti-correlated
patterns in correlated two-electron momentum distributions, and
also the transition between these two patterns. Being
significantly enhanced relatively by the recapture process,
recolliding trajectories of the second electron excited by the
first- or third-return recolliding trajectories of the first
electron produce the cross-shaped or anti-correlated
distributions, respectively. And the transition is induced by the
increasing contribution of the third return with increasing pulse
duration. Our work provides new insight into atomic ionization
dynamics and paves the new way to imaging of ultrafast dynamics of
atoms and molecules in intense laser field.
\end{abstract}

\maketitle

Recollision is responsible for many intriguing strong-field
phenomena, such as high-order above-threshold ionization (HATI),
high harmonics generation (HHG), and nonsequential double
ionization (NSDI), and also serves as the foundation of attosecond
physics (see, e.g., Refs.
\cite{Proto1997,Becker2002,Becker2012,Krausz2009} for reviews and
references therein). In the recollision picture
\cite{Corkum1993,Schafer1993}, an electron is liberated from the
neutral atom or molecule through tunneling, then is driven back by
the laser field to collide with the parent ion elastically or
inelastically, or recombine with the ion, resulting in HATI, NSDI
and HHG, respectively. Since the electron strongly interacts with
the ion, the products upon recollision carry information of the
parent ion, and can be used to probe its structure and dynamics.
Based on the recollision process, different methods, such as
laser-induced electron diffraction (LIED) \cite{Morishita2008PRL}
and laser-induced electron inelastic diffraction (LIID)
\cite{Quan2017}, are proposed and successfully applied in imaging
of atomic and molecular ultrafast dynamics and structure with
unprecedented spatial-temporal resolution
\cite{Okunishi2008PRL,Ray2008PRL,Meckel2008,Blaga2012,Niikura2002,Itatani2004,Wolter2016Science,Quan2017}.
However, the recollision in the above-mentioned strong-field
processes and ultrafast imaging methods is limited to interaction
between the ionized electron and univalent ion.

\begin{figure}[ptb]
\begin{center}
\includegraphics[width=3in]{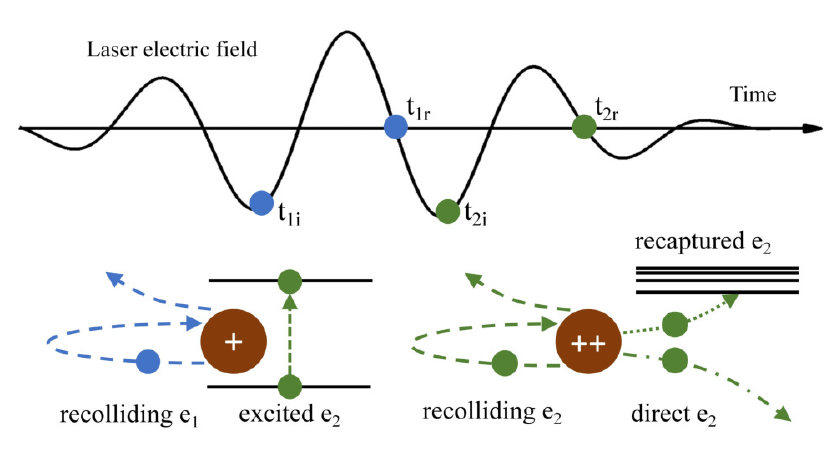}
\end{center}
\caption{Sketch map to illustrate the below-threshold NSDI
process. At time $t_{1i}$, $e_1$ is first ionized by the laser
field, then it is driven back to collide the parent univalent ion
and excites $e_2$ at time $t_{1r}$. $e_2$ is ionized from the
excited state by the laser field at a later time $t_{2i}$. After
that, $e_2$ may travel directly to the detector, or it may be
driven back to recollide with the divalent ion similar to $e_1$,
or it may also be recaptured into a Rydberg state of ion.}
\label{fig1}%
\end{figure}

In the NSDI process, one electron ($e_1$) firstly experiences a
recollision with the parent univalent ion and deliver energy to
the bounded electron ($e_2$). In the below-threshold regime, the
maximal kinetic energy of $e_1$ upon recollision is smaller than
the ionization potential of $e_2$, so $e_2$ can be only pumped to
an excited state, as illustrated in Fig. \ref{fig1}. Then $e_2$ is
ionized from the excited state by the laser field at a later time,
dubbed as recollision excitation with subsequent ionization (RESI)
process. Usually, it is believed that $e_2$ will travel directly
to the detector \cite{Faria2010,Wang2012,Hao2014,Maxwell2016}.
However, after tunneling ionization, $e_2$ may be driven back to
recollide with the divalent ion or be recaptured into a Rydberg
state of ion as illustrated in Fig. \ref{fig1}. Due to the strong
Coulomb field of the divalent ion, these post-tunneling dynamics
may be prominent. It has been recently reported experimentally and
theoretically that the probability of recapture in double
ionization, dubbed as frustrated double ionization (FDI), is much
higher than expectation \cite{Larimian2020,Chenshi2020}.

In this work, by introducing a Coulomb-corrected
quantum-trajectories (CCQT) method, we identify the key role
played by the recollision between the second ionized electron and
the divalent ion in the below-threshold NSDI process. We find
that, only when this recollision is included, the experimentally
observed cross-shaped \cite{Bergues2012,Kubel2016} and
anti-correlated \cite{Liu2008} patterns of correlated electron
momentum distribution (CEMD), and also the transition between them
\cite{Kubel2014}, can be well reproduced.

To describe the below-threshold NDSI process both coherently and
quantitatively, it has to incorporate both the quantum effect and
the Coulomb interaction between the residual ion and the ionized
electrons in a uniform theory. To achieve this, we introduce a
Coulomb-corrected quantum-trajectories (CCQT) method by taking
advantage of the well-developed Coulomb-corrected methods dealing
with single-electron dynamics. The transition magnitude
is expressed as (atomic units $m=\hbar=e=1$ are used)%
\begin{align}
&  M\left(  \widetilde{\mathbf{p}}_{1},\widetilde{\mathbf{p}%
}_{2}\right)  \nonumber\\
&  =%
{\displaystyle\sum\limits_{s}}
M_{\widetilde{\mathbf{p}}_{2}}^{\left(  3\right)  }\left(  t_{2i}^{s},t_{1r}%
^{s}\right)  M_{\widetilde{\mathbf{p}}_{1}}^{\left(  2\right)
}\left( t_{1r}^{s}\right)  M_{\widetilde{\mathbf{p}}_{1}}^{\left(
1\right)
}\left(  t_{1r}^{s},t_{1i}^{s}\right),  \label{m}%
\end{align}
in which different trajectories labelled with $s$ are summed
coherently. $M_{\widetilde{\mathbf{p}}_{1}}^{\left( 1\right)
}\left( t_{1r}^{s},t_{1i}^{s}\right)  $, describing the tunneling
ionization of $e_1$ at $t_{1i}^{s}$\ and its subsequent
propagation in the laser field until colliding with the parent ion
at time $t_{1r}^{s}$, is calculated using the quantum-trajectory
Monte Carlo (QTMC) method \cite{Li2014,Song2016} which is
efficient to obtain large amount of hard-collision trajectories.
Trajectories with minimum distance from the ion less than 1 a.u.
are selected to consider the hard collision for the subsequent
calculation. Upon collision, $e_1$ will excite $e_2$ and then move
to the detector. This excitation process is described by
$M_{\widetilde{\mathbf{p}}_{1}}^{\left( 2\right) }\left(
t_{1r}^{s}\right)  $ which is calculated with conventional
S-matrix theory. Finally, $e_2$ is ionized through tunneling at
$t_{2i}^{s}$ from the excited state, and then propagates in the
laser field until the end of the pulse, which is described by
$M_{\widetilde{\mathbf{p}}_{2}}^{\left(  3\right)  }\left(  t_{2i}^{s}%
,t_{1r}^{s}\right)  $ calculated with the Coulomb-corrected strong
field approximation (CCSFA) method \cite{Yan2010}. The sin-squared
pulse shape is employed in our calculation. A model potential
\cite{Tong2005} is applied to mimic the Coulomb field of Ar$^{2+}$
felt by $e_2$ in its propagation. Only the first excited state
$3s3p^{6}$ with zero magnetic quantum number \cite{grasp} is
included in the present calculations. The
depletion of the excited state is also taken into account in calculating $M_{\widetilde{\mathbf{p}}_{2}}^{\left(  3\right)  }\left(  t_{s}%
,t_{s}^{\prime}\right)  $ \cite{Hao2014} (see the method in
Supplement 1).

\begin{figure}[pt]
\centering\vspace{-0in}\includegraphics[width=3.3in]{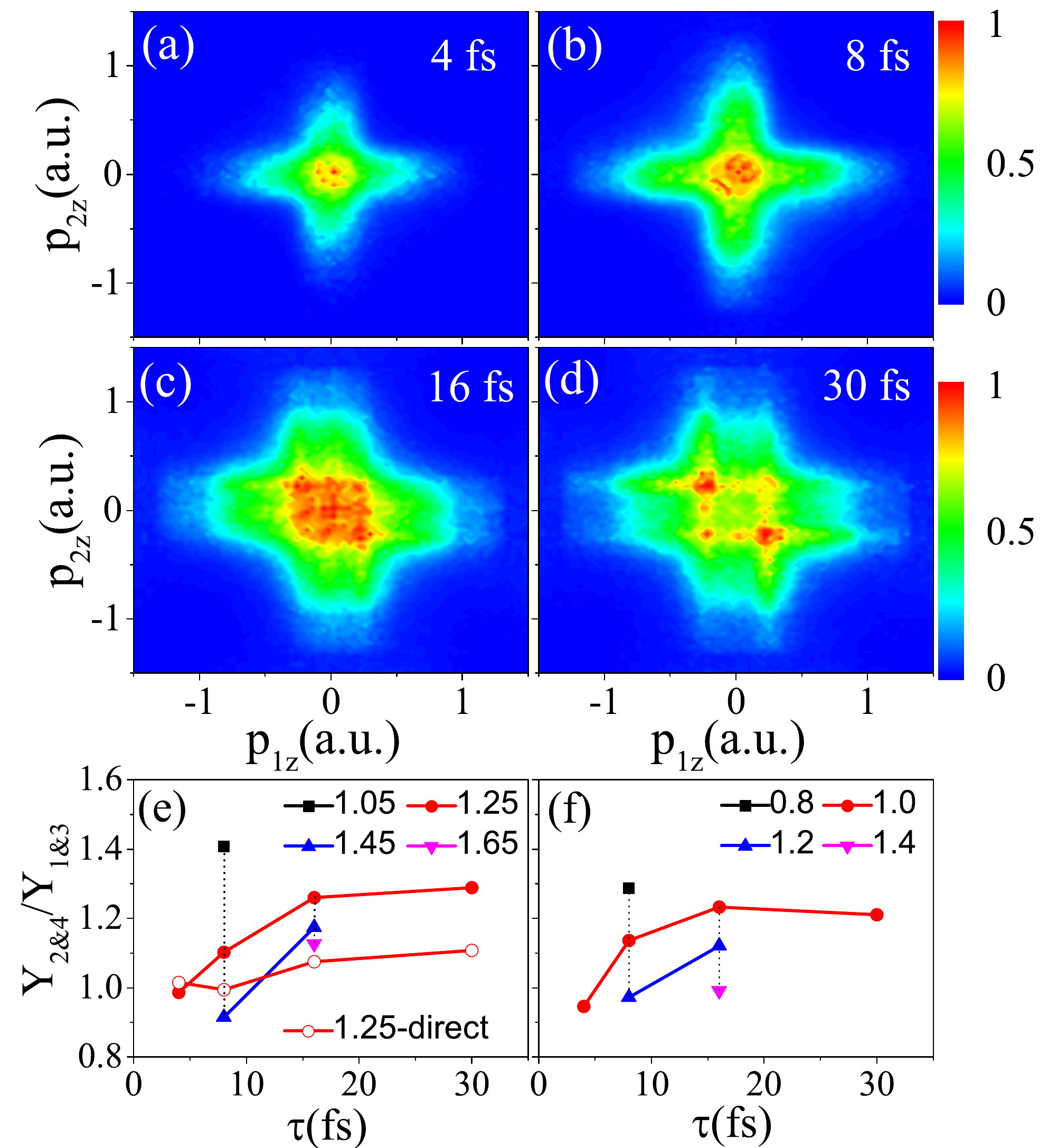}~\vspace
{-0.1in}~~\caption{(a)-(d) Simulated CEMDs of Ar for different
laser pulse durations at the intensity of $1.25\times10^{14}$
W/cm$^{2}$. The CEPs are averaged. Each CEMD is normalized to
itself. (e) Simulated yield ratio $Y_{2\&4}/Y_{1\&3}$ for
different pulse durations and different intensities. $Y_{1\&3}$
($Y_{2\&4}$) denotes the integrated yield in the first and third
(the second and fourth) quadrants in the CEMD. The numbers given
in the legends denote peak laser intensities with units of
$10^{14}$ W/cm$^{2}$. The open circles are calculated by only
considering direct trajectories of $e_2$ (see text for details).
(f) Measured results extracted from Ref.\cite{Kubel2014}. The
black short-dashed lines in (e) and (f)
serve as indications of the intensity dependence.}%
\label{fig2}%
\end{figure}

Fig. \ref{fig2} displays the calculated results for Ar under
different pulse durations to compare with the experimental results
in Ref. \cite{Kubel2014}. Intensities higher than the measured
ones by $0.25\times10^{14}$ W/cm$^{2}$ are used in the present
calculations (see Supplement 1 for details of the fitting
procedure). As shown in Fig. \ref{fig2}, for shorter pulse
durations (4 fs and 8 fs), the distributions show a cross shape
with the maxima lying at the origin. While for longer pulses (16
fs and 30 fs), the electrons are more homogeneously distributed
over the four quadrants, actually, prefer the second and fourth
quadrants, which indicates an anti-correlation. This transition of
CEMD from cross-shaped to anti-correlated patterns is in agreement
with the measured results reported in Ref. \cite{Kubel2014},
although there is some discrepancy in details. In the measurement,
the transition occurs when pulse duration increases from 4 fs to 8
fs, whereas in Fig. \ref{fig2} it occurs when pulse duration
increases from 8 fs to 16 fs. This discrepancy may be due to that
the pulse shape and duration employed in our calculations are not
exactly the same as that in the measurements.

To quantitatively characterize the CEMD, in Fig. \ref{fig2}(e) we
plot the ratio $Y_{2\&4}/Y_{1\&3}$ for different pulse durations
and different intensities. $Y_{1\&3}$ ($Y_{2\&4}$) denotes the
integrated yield in the first and third (the second and fourth)
quadrants. We also present the measured results \cite{Kubel2014}
in Fig. \ref{fig2}(f) for comparison. In general, the simulation
reproduces most of the features in the measured results. The ratio
increases with pulse duration and becomes saturated at 16 fs when
the intensity is fixed, and it decreases with laser intensity both
for pulse durations of 8 fs and 16 fs. However, compared with the
measured results, the simulation obviously overestimates the ratio
for the highest intensity. This discrepancy can be attributed to
that the contribution of the process that $e_2$ is directly
knocked out by $e_1$, whose distribution mainly locates in the
first and third quadrants, becomes more significant with
increasing intensity, but is not included here.

\begin{figure}[pt]
\centering\vspace{-0in}\includegraphics[width=3.3in]{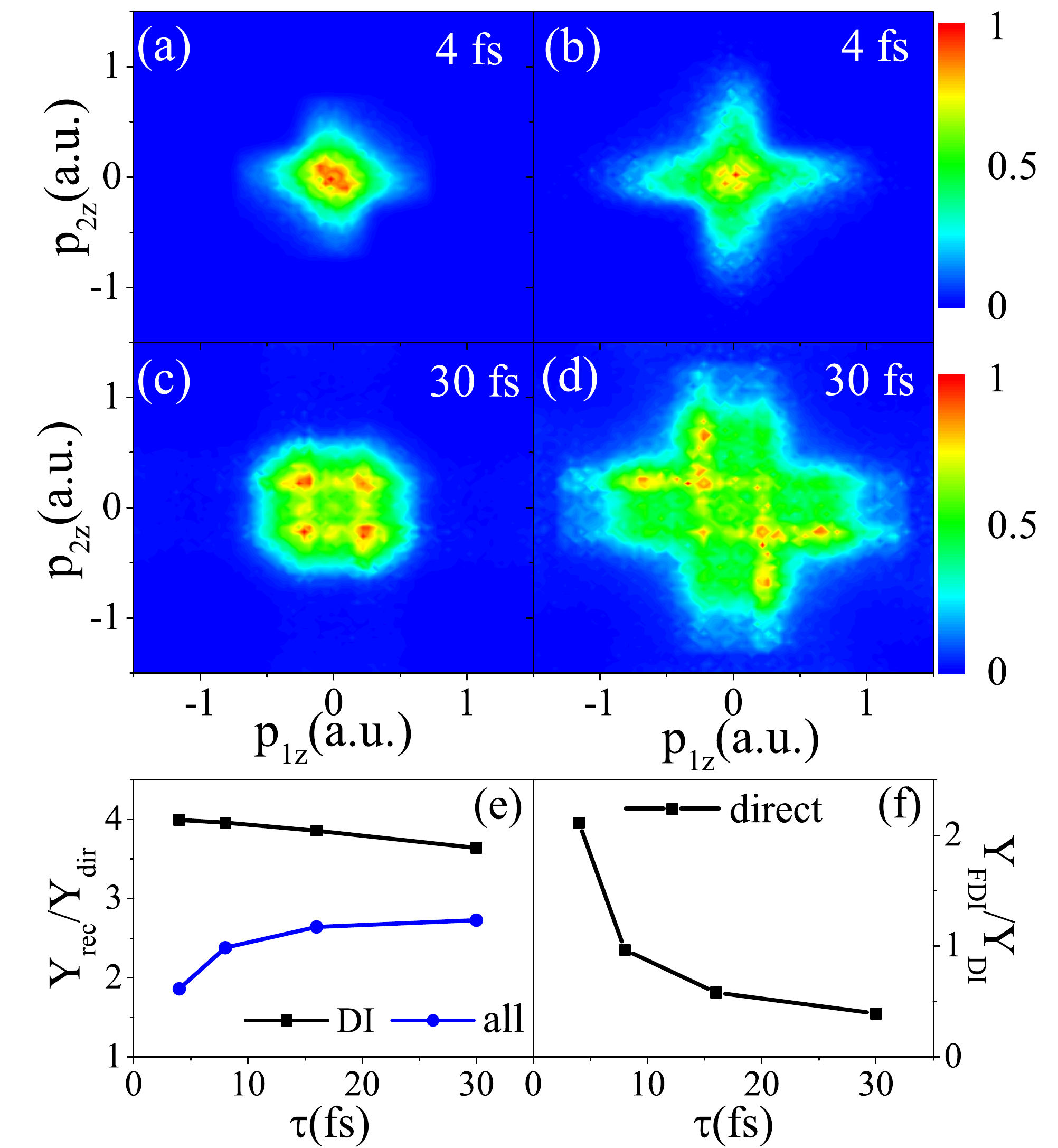}~\vspace
{-0.1in}~~\caption{CEMDs corresponding to direct trajectories [(a)
and (c)] and recolliding trajectories [(b) and (d)] of $e_2$. Each
CEMD is normalized to itself. (e) Pulse-duration dependence of
$Y_{rec}/Y_{dir}$, the ratio between the integrated yields of
recolliding and direct trajectories for $e_2$ for all events or
only double ionization (DI) events. (f) Pulse-duration dependence
of $Y_{FDI}/Y_{DI}$, the ratio between the probabilities of FDI
and DI when $e_2$ is confined to direct trajectories.
The laser intensity is $1.25\times10^{14}$ W/cm$^{2}$.}%
\label{e2-FDI}%
\end{figure}

In Fig. \ref{e2-FDI}, we present CEMDs corresponding to
recolliding trajectories and direct trajectories of $e_2$ at 4 fs
and 30 fs, respectively. Here, we define it as the recolliding
trajectory if the minimal distance of $e_2$ from the residual ion
is less than the tunnel exit. Otherwise, it is the direct
trajectory. Since momenta of direct trajectories of $e_2$ are much
smaller than that of recolliding trajectories, CEMDs for direct
trajectories are localized around the origin for both 4 fs and 30
fs pulses, as shown in Figs. \ref{e2-FDI}(a) and \ref{e2-FDI}(c).
Whereas the CEMD for recolliding trajectories exhibits a cross
structure at 4 fs [Fig. \ref{e2-FDI}(b)], and exhibits an
anti-correlated pattern at 30 fs [Fig. \ref{e2-FDI}(d)].
Meanwhile, recolliding trajectories of $e_2$ have dominant
contributions for all pulse durations as depicted by the ratio
$Y_{rec}/Y_{dir}$ ($Y_{rec}$ and $Y_{dir}$ denote the yields of
recolliding and direct trajectories, respectively) for double
ionization (DI) events in Fig. \ref{e2-FDI}(e), as a consequence,
the total CEMDs also shows a cross or an anti-correlated pattern
at 4 fs or 30 fs, respectively.

But why the relative contribution of the recolliding trajectories
of $e_2$ is so high? Intuitively, the Coulomb focusing effect
imposed on $e_2$ by the divalent cation, which is much stronger
than that of the univalent cation in ATI process, will effectively
enhance the probability of recollision. We can indeed see this
clearly from Fig. \ref{e2-FDI}(e) in which the ratio
$Y_{rec}/Y_{dir}$ with all events included is greater than 1. But
it is still much smaller than the ratio considering only DI
events. This deviation is the result of the important contribution
of recapture or FDI process. More than two-thirds of direct $e_2$
are recaptured into the Rydberg states of Ar$^+$ at 4fs, and the
probability of FDI for direct $e_2$ decreases quickly with
increasing pulse duration, as shown in Fig. \ref{e2-FDI}(f).
Compared with recolliding trajectory of $e_2$, direct $e_2$ cannot
move far away from Ar$^{2+}$ at the end of the pulse due to its
much lower momentum, especially in shorter laser pulse, therefore
is easier to be recaptured by the strong Coulomb field of the
divalent ion. More direct $e_2$ being recaptured means fewer of
them contribute to DI, resulting in larger relative contribution
of recolliding trajectories of $e_2$ to DI. In brief, the enhanced
FDI probability significantly enlarges the relative contribution
of recolliding trajectories of $e_2$ to DI, and eventually induces
the experimentally observed cross-shaped and anti-correlated
patterns. In addition, this point is strongly supported by the
fact that when only the direct trajectories of $e_2$ are
considered, the calculated $Y_{2\&4}/Y_{1\&3}$ is significantly
different from the experimental result [see Fig. \ref{fig2}(e)].

\begin{figure}[pt]
\centering\vspace{0in}\includegraphics[width=3.3in]{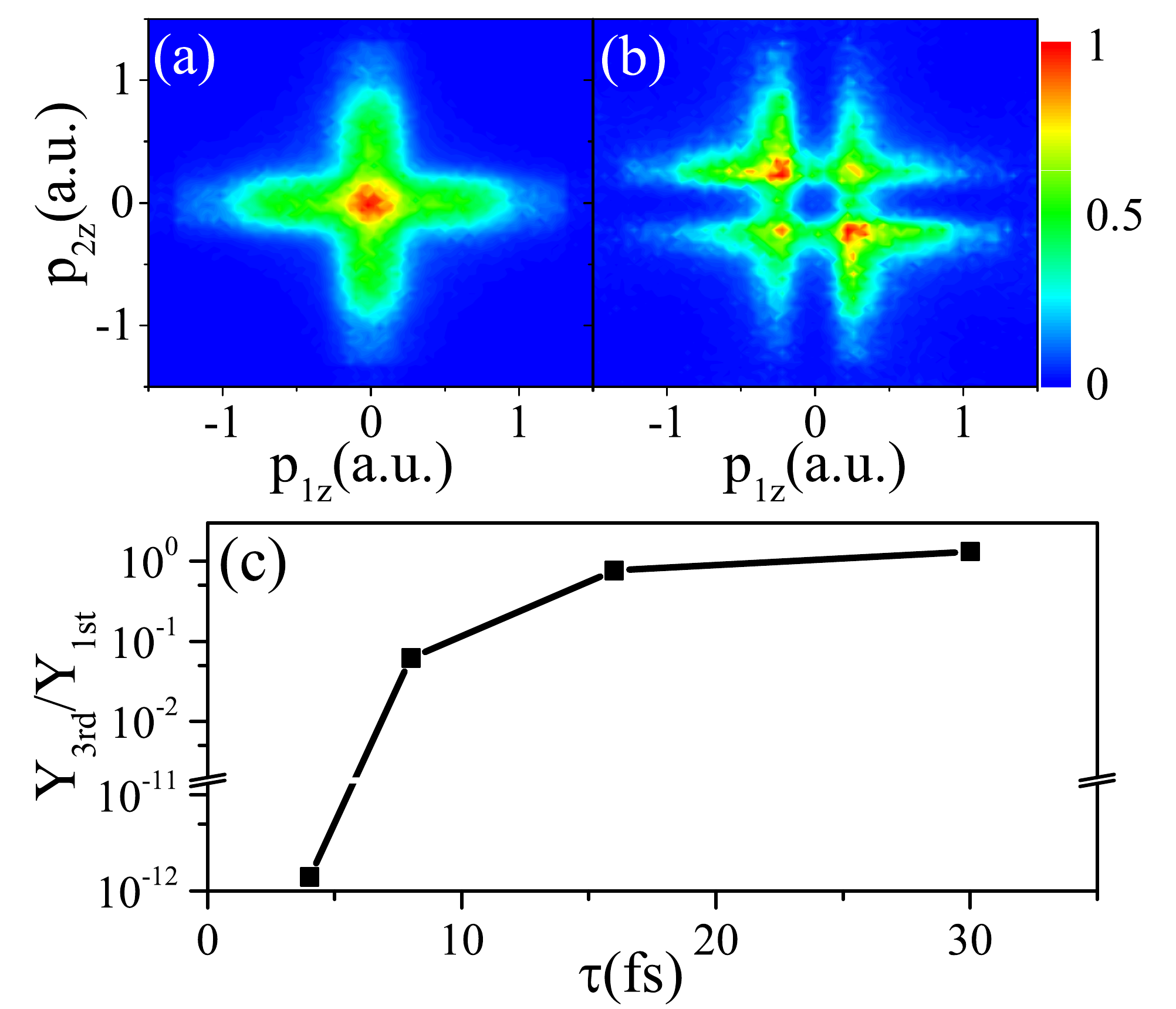}~\vspace
{0.2in}\caption{CEMDs in 30 fs, $1.25\times10^{14}$ W/cm$^{2}$
laser pulse when the trajectories of $e_1$ are confined to the
first return (a) or the third return (b). Each CEMD is normalized
to itself. (c) Pulse-duration dependence of $Y_{3rd}/Y_{1st}$, the
ratio of the integrated yield of the third-return to that of the
first-return trajectories of
$e_1$.}%
\label{e1-13}%
\end{figure}

The specific pattern of CEMD also requires the appropriate
momentum of $e_1$ which is determined by the microscopic dynamics
of the recollision process for $e_1$. In the recollision process
of $e_1$, it may miss the parent ion at its first return but
collide with the ion at the subsequent returns. In our model, the
different-return trajectories of $e_1$ can be distinguished
according to the travel time $t_{t}$ defined as the interval
between the ionization time $t_{1i}$ and the recollision time
$t_{1r}$. For trajectories with $t_{t}$ in the interval
[$(n/2)T,~((n+1)/2)T$] ($T$ is the optical cycle), we denote them
as the $n$th-return trajectories \cite{HaoPRA2011,HaoPRA2020}.
According to our calculations, the first- and third-return
recolliding trajectories of $e_1$ are dominant for the laser
parameters interested here. For other returns, either the return
energy is too small to excite $e_2$, or the collision probability
is negligible due to the spreading of the wave packet. In Figs.
\ref{e1-13}(a) and \ref{e1-13}(b), we present the CEMDs
corresponding to the first- and third-return trajectories of
$e_1$, respectively, in $1.25\times10^{14}$ W/cm$^{2}$, 30 fs
laser pulse. Note that all trajectories of $e_2$ (direct and
recolliding trajectories) are included. The CEMD for the
first-return trajectories of $e_1$ [Fig. \ref{e1-13}(a)] shows a
cross-shaped pattern, whereas that for the third-return
trajectories [Fig. \ref{e1-13}(b)] exhibits an anti-correlated
pattern. As shown in Fig. \ref{e1-13}(c), the ratio of the
integrated yield of the third-return trajectories to that of the
first-return increases quickly with increasing pulse duration.
Correspondingly, the CEMD changes from a cross-shaped to an
anti-correlated pattern. Therefore, the transition between the two
patterns of CEMD with increasing pulse duration is the result of
increasing contribution of the third-return trajectories of $e_1$.
The significant contribution of the third-return trajectories can
be attributed to the Coulomb focusing effect from the univalent
cation. The similar effect has also been reported for high-order
ATI process \cite{HaoPRA2020}.

\begin{figure}[pt]
\centering\vspace{-0in}\includegraphics[width=3.3in]{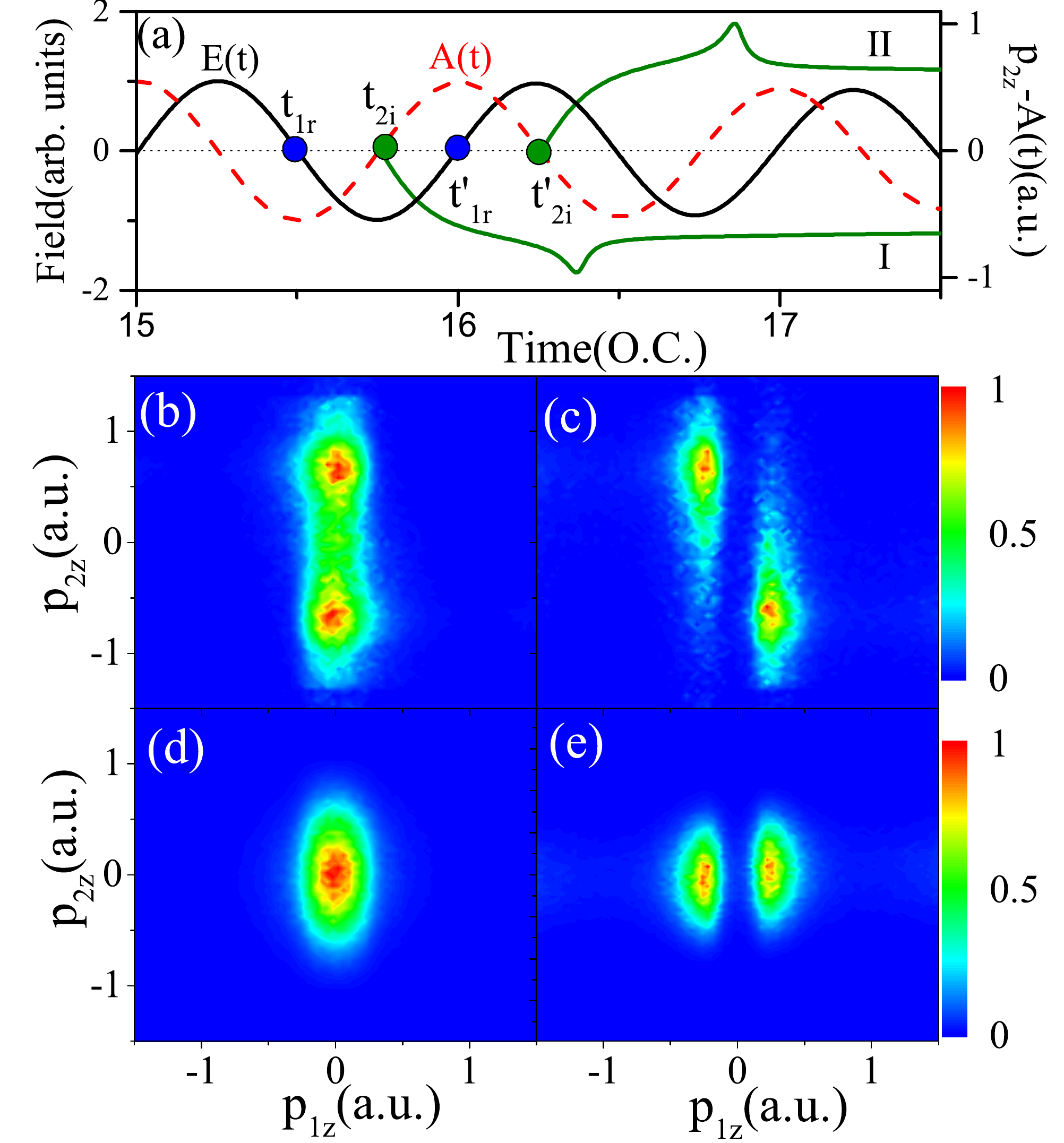}~\vspace
{0.2in}\caption{(a) Schematic representation of the laser electric
field $\mathbf{E}\left(  t\right)$ and the corresponding vector
potential $\mathbf{A}\left( t\right) $ for pulse duration of 30
fs. $e_1$ collides with the ion most probably at the crossing of
$\mathbf{E}\left( t\right)$ at $t_{1r}$ or $t_{1r}^{\prime}$. Upon
the collision, $e_2$ is excited, and then is ionized most probably
at the peak of the laser field at $t_{2i}$ or $t_{2i}^{\prime}$.
The subsequent evolution of the canonical momentum $p_{2z} -
A\left( t\right)$ for the recolliding trajectories of $e_2$,
denoted as I and II, are presented to illustrate the Coulomb-field
effect of $Ar^{2+}$. (b) and (c) CEMDs without performing electron
indistinguishability symmetrization, and only recolliding
trajectories of $e_2$ with ionization time nearest the collision
time of $e_1$ are included. (d) and (e) CEMDs calculated by
replacing the ionization amplitude
$M_{\widetilde{\mathbf{p}}_{2}}^{\left( 3\right) }$ in Eq.
(\ref{m}) with the standard SFA. Trajectories of $e_1$ are
confined to the first return in (b) and (d),
and the third return in (c) and (e). Each CEMD is normalized to itself.}%
\label{e2-coulomb}%
\end{figure}

Next, we will explain how the cross-shaped and anti-correlated
patterns of CEMDs are formed by the recolliding trajectories of
the two electrons. Without indistinguishability symmetrization,
the first-return trajectories of $e_1$ will show a band-like
distribution along the $p_{1z}=0$ axis with the maxima away from
the origin, i. e., vanishing momentum of $e_1$ but much higher
momentum of $e_2$ [Fig. \ref{e2-coulomb}(b)]. Whereas the CEMD for
the third-return consists of two bands and the maximum of the left
(right) band lies in the up (low) part, giving rise to an
anti-correlation [Fig. \ref{e2-coulomb}(c)]. These band-like
distributions can be understood as follows. The final momentum of
$e_1$ is determined by the residual momentum after exciting $e_2$
and the drift momentum it obtains from the laser field. Since
forward scattering is favored in this inelastic scattering
process, the residual momentum and the drift momentum are in
opposite directions and will cancel with each other. At the
present intensity ($1.25\times10^{14}$ W/cm$^{2}$), the magnitudes
of them for the first-return trajectories of $e_1$ are nearly
equal, resulting in a vanishing momentum of $e_1$. When the laser
intensity increases, the band will become tilted towards the main
diagonal \cite{Kubel2016} due to the faster-increasing residual
momentum. For the third-return, its return energy is smaller than
that of the first-return, so the residual momentum is not enough
to compensate the drift momentum, resulting in a non-vanishing
momentum of $e_1$. Since electrons ionized at times separated by a
half optical cycle will have opposite momenta, there is one band
on each side of $p_{1z}=0$ axis. Actually, there are also two
bands for the first return, but they merge together.

The anti-correlation between the two electrons for the
third-return trajectories of $e_1$ is illustrated in Fig.
\ref{e2-coulomb}(a). The recollision of $e_1$ most probably occurs
around the crossing of the electric field at $t_{1r}$ or
$t_{1r}^{\prime}$. Since the magnitude of the drift momentum after
recollision, which is equal to $-\mathbf{A}\left( t_{r}\right) $
(vector potential at the recollision time), is larger than the
residual momentum for the third-return recolliding trajectories of
$e_1$, its final momentum is in the direction of the drift
momentum. If the recollision of $e_1$ occurs at $t_{1r}$, the
final momentum of $e_1$ will be positive, corresponding to the
right band in Fig. \ref{e2-coulomb}(c). Upon recollision, $e_2$ is
pumped to the first excited state, then it is most probably
ionized at the subsequent electric field peak at $t_{2i}$. If the
Coulomb attraction of the ion is not considered and no recollision
occurs, $e_2$ will have vanishing final momentum. This can be seen
clearly in Figs. \ref{e2-coulomb}(d) and \ref{e2-coulomb}(e), in
which the CEMDs are obtained by calculating $M_{\widetilde
{\mathbf{p}}_{2}}^{\left( 3\right) }$ in Eq. (\ref{m}) with the
standard SFA. But if the ionic Coulomb potential is taken into
account, momenta of $e_2$ for recolliding trajectories (trajectory
I) shift to the negative direction, opposite to the direction of
the final momentum of $e_1$ [see Fig. \ref{e2-coulomb}(a)]. This
is exactly the situation of the right-band distribution in Fig.
\ref{e2-coulomb}(c). The left band corresponds to the situation
that $e_1$ recollides with the ion at $t_{1r}^{\prime}$ and $e_2$
is ionized at $t_{2i}^{\prime}$. As a consequence, the two
electrons are emitted back-to back and the CEMD exhibits an
anti-correlated pattern. In addition, it is also possible that the
recollision of $e_1$ occurs at $t_{1r}$ while $e_2$ is ionized at
$t_{2i}^{\prime}$, which will produce a correlated CEMD. But since
its contribution is smaller due to the depletion effect of the
excited state, the total CEMD will still exhibit an
anti-correlated pattern.

In conclusion, we propose a Coulomb-corrected quantum-trajectories
(CCQT) method to describe the below-threshold NSDI process both
coherently and quantitatively. It enables us to well reproduce
different kinds of CEMDs observed in experiments, and uncover the
rich underlying physics which is enhanced by the Coulomb field of
univalent and divalent ions, including the multi-return
trajectories of the first ionized electron $e_1$, the recollision
and recapture processes of the second ionized electron $e_2$.
Especially, recollision process of $e_2$, which is enhanced
relatively by the recapture process of $e_2$, is found to play an
important role in electron-electron correlation. We expect that
the recollision process of $e_2$ can be applied to develop a new
scheme to image the ultrafast evolution of the molecular structure
and dynamics induced by the strong laser field.

This work was partially supported by the National Key Program for
S\&T Research and Development (No. 2019YFA0307700 and No.
2016YFA0401100), the National Natural Science Foundation of China
(Grants No. 11874246, No. 91950101, No. 11774215).

\end{document}